\documentclass[english,11pt,reqno]{smfart}

\usepackage{amssymb}
\usepackage{amsmath}
\usepackage{amsthm}
\usepackage{dsfont}
\usepackage{graphicx}
\usepackage{appendix}
\usepackage{graphicx}
\usepackage{a4wide}
\usepackage{wasysym}
\usepackage{textcomp}

\newtheorem{theorem}{Theorem}

\newtheorem{lemma}[theorem]{Lemma}
\newtheorem{corollary}[theorem]{Corollary}

\newtheorem{remark}{Remark}

\newtheorem{assumption}{Assumption}

\def\di{\displaystyle}

\newcommand{\R}{\mathbb{R}}

\newcommand{\grad}[1]{\mathsf{grad}\, #1}
\renewcommand{\div}[1]{\mathsf{div}\, #1}
\newcommand{\lap}[1]{\mathsf{lap}\, #1}
\newcommand{\Boxinf}{\Box_\infty}
\newcommand{\X}{\textbf{X}}
\newcommand{\V}{\textbf{V}}

\begin{document}
\setcounter{tocdepth}{3}
\title[Scale dynamical origin of modification or addition of potential]{Scale dynamical origin of modification or addition of potential in mechanics. A possible framework for the MOND theory and the dark matter.}
\author{Fr\'ed\'eric Pierret}
\address{IMCCE, Observatoire de Paris, France}

\begin{abstract} 
Using our mathematical framework developed in \cite{cresson-pierret_scale} called \emph{scale dynamics}, we propose in this paper a new way of interpreting the problem of adding or modifying potentials in mechanics and specifically in galactic dynamics. An application is done for the two-body problem with a Keplerian potential showing that the velocity of the orbiting body is constant. This would explain the observed phenomenon in the flat rotation curves of galaxies without adding \emph{dark matter} or modifying Newton's law of dynamics. 
\end{abstract}

\maketitle

\tableofcontents

\section{Introduction}
Nowadays, one of the most important problem in galactic dynamics and cosmology concerns the so-called \emph{missing mass} or \emph{dark matter} (see \cite{sanders2002_2}). In particular, this mass would be responsible of the flat rotation curves of spiral galaxies. The usual approach is to add a huge amount of invisible and unknown matter to solve the problem. However, having direct observational evidences of the presence and proof of the existence of such kind of matter is still a failure. Another approach has been proposed in \cite{milgrom1983_5} known as MOND for MOdified Newtonian Dynamics. It consists in modifying Newton's equation which is a linear relation between the force and the acceleration in a non-linear one when the acceleration is weak with respect to a constant fixed by the theory. Up to a choice of functions in the theory, it well explains at the galactic level the behavior of the rotation curves (see e.g. \cite{bekenstein1984,milgrom2009,bla2011} with references therein for a review of the best formulation of MOND at the non-relativistic level with a modified Poisson equation). In both cases, it can be seen as adding a potential of matter or modifying the initial potential. \\

In our recent work \cite{cresson-pierret_scale}, we developed a mathematical formalism to take into account \emph{scale formulation of dynamical systems}. When modeling a dynamical system, the notion of \emph{scale} naturally occurs at different levels. For example, taking into account the notion of scale can be of dynamical origin as in fluid dynamics or geometric as in the study of certain fractal objects. More generally, this concept appears when attempting to characterize the nature of a trajectory. Indeed, for example in mechanics, a finite set of observational data is limited by a scale resolution. From this problem, we developed the formalism of \emph{scale dynamics} which allows taking into account dynamical effects induced by the nature of the observed behavior. To illustrate this formalism, we have shown that the Newton equation and the Schr\"odinger equation were equivalent assuming that the nature of the trajectories was fractional. This is the case for typical trajectories in quantum mechanics (see \cite{feynman}). This approach is different from the usual one in classical mechanics. Indeed, from our point of view, classical mechanics is formulated assuming a linear scale regime, i.e. the asymptotic models has a differentiable motion. Such an approach to change the scale regime can be used to think about a new possible explanation of the problem of dark matter, and by consequent, to the MOND theory, using scale dynamics. This idea is supported by the work of L. Nottale on \emph{scale relativity}. Indeed, he used the notion of \emph{fractal space-time} to show that the constant velocity curves in galaxies can be explained by the \emph{fractal behavior} of the dynamics (see \cite{nottale1993fractal,nottale2011} and references therein). In this paper, we propose to apply our formalism to mechanics to show that the scale formulation of the Newton equation exhibit dynamical effects induced by the scale regime observed or chosen. These supplementary dynamical terms can be interpret as a modification or an addition of potentials. Moreover, applying it to the two body problem, with a particular fractional scale regime, we recover the result obtained by L. Nottale, i.e. the velocity of the orbiting body is constant.\\

The plan of the paper is as follows:\\

In Section \ref{reminder}, we remind the basic tools from scale dynamics concerning the asymptotic formulation of models and operators under a particular scale regime. In Section \ref{asymptotic}, we develop the main equations of this paper from the scale formulation of Newton's equation. Particularly, under a fractional scale regime, we obtain a fractional Hamilton-Jacobi equation and a fractional version of the Virial theorem. In Section \ref{kepler}, we apply the development in the previous section on the two-body problems and we show that the velocity of the orbiting body tend to be constant.\\

\section{Reminder about asymptotic models and differential operators from scale dynamics}
\label{reminder}

We remind in this section the basis of scale dynamics about the asymptotic models and operators constructed from the scale formulation and definitions of objects. The definition are directly extended to $\R^d$. For complete details, we refer to our paper \cite{cresson-pierret_scale}. \\

The construction of asymptotic models and differential operators are based from the global behavior observed of a phenomenon at different scale, i.e. the observed scale regime. From this, one can assume that on this scale range, the asymptotic behavior will be the same. Doing so, one can obtain the construction of an asymptotic model representing the phenomenon at the continuous level in a \emph{regular part} and a \emph{deviant part} which contains the information on the observed behavior. For example, observing a linear scale regime on a specific scale range, one can construct an asymptotic continuous model which will be, from the definition of a linear scale regime, differentiable. In consequence, the deviant part will be zero. But, observing a fractional scale regime, one can construct an asymptotic model which will have by definition, a non zero deviant part. For any asymptotic model $\X_\infty\in\R^d$, we denote
\begin{equation}
\X_\infty = \X_\infty^\star + \textbf{D}_\infty
\end{equation}
where $\X_\infty^\star$ is the regular part of $\X_\infty$ and $\textbf{D}_\infty$ is the deviant part of $\X_\infty$. The $\Delta_{\infty}$ and $\nabla_{\infty}$ derivatives are respectively defined for any asymptotic models $\X_\infty$ as
\begin{equation}
\Delta_{\infty} \X_\infty :=\frac{d^+}{dt} \X_\infty^{\star} \quad \text{and} \quad \nabla_{\infty} \X_\infty :=\frac{d^-}{dt} \X_\infty^{\star}
\end{equation}
In fact, the operators $\Delta_{\infty}$ and $\nabla_{\infty}$ extract the right and left derivatives of the regular part of $\X_{\infty}$. We are interested in asymptotic models which are not differentiable, i.e. asymptotic models which are not obtained with a linear scale regime. The ``first simple'' class of such models is characterized by the comparison class
\begin{equation}
\mathcal{F} =\left \{ f_{\alpha} (t)=t^{\alpha},\ 0<\alpha<1  \right \}.
\end{equation}
The class $\mathcal{F}$ defines the fractional scale behavior. We assume that $\X_\infty$ has a fractional scale regime of order $\alpha$ on each its dimension, and we denote by $j_{\alpha} =E(1/\alpha)$, the integer part of $1/\alpha$. The definitions of $\Delta_{\infty}$ and $\nabla_{\infty}$ derivatives over sufficiently smooth function $f(t,\X)$ are given by

\begin{equation}
\Delta_{\infty}  f(t,\X_{\infty}) : = \di\frac{d^+}{dt} f(t,\X_{\infty}^{\star} ) + \sum_{k_1,\dots,k_{j_\alpha}=1}^{d}\frac{\lambda^+_{k_1} \dots \lambda^+_{k_{j_\alpha}}}{j_\alpha!}\,\frac{\partial^{j_\alpha} f(t, \X_{\infty}^{\star} )}{\partial k_1 \dots \partial k_{j_\alpha}}
\end{equation}
and
\begin{equation}
\nabla_{\infty}  f(t,\X_{\infty}) : = \di\frac{d^-}{dt} f(t,\X_{\infty}^{\star} ) +(-1)^{j_\alpha-1} \sum_{k_1,\dots,k_{j_\alpha}=1}^{d}\frac{\lambda^-_{k_1} \dots \lambda^-_{k_{j_\alpha}}}{j_\alpha!}\,\frac{\partial^{j_\alpha} f(t, \X_{\infty}^{\star} )}{\partial k_1 \dots \partial k_{j_\alpha}}
\end{equation}
where $\lambda^+=\left(\lambda^+_1, \dots, \lambda^+_d \right)^\mathsf{T}$ and $\lambda^-=\left(\lambda^-_1, \dots, \lambda^-_d \right)^\mathsf{T}$ are the vectors containing the comparison constants for the $\Delta_{\infty}$ and $\nabla_{\infty}$ derivatives of the fractional regime on each dimension of $\R^d$. In order to compute quantities containing all the dynamical information contained in the $\Delta_\infty$ and $\nabla_\infty$ derivatives, one can consider the differential operator denoted by $\Box_{\infty}$ which is the linear operator defined by
\begin{equation}
\Box_{\infty} = \di\frac{1}{2} \left (\Delta_{\infty} +\nabla_{\infty} \right ) + i\frac{\eta }{2} \left ( 
\Delta_{\infty} -\nabla_{\infty} \right ) ,  
\end{equation}
where $i^2 =-1$ and $\eta =\{ -1,1,-i,i \}$. In that case, we have
\begin{equation}
\Box_\infty  f(t,\X_{\infty}) := \frac{\Box}{\Box t}  f(t,\X^\star_{\infty})+   \frac{\lambda_{k_1,\dots,k_{j_\alpha}}}{j_{\alpha} !} \frac{\partial^{j_\alpha} f(t, \X_{\infty}^{\star} )}{\partial k_1 \dots \partial k_{j_\alpha}},
\end{equation}
where
\begin{equation}
	\frac{\Box}{\Box t}:=\frac{1}{2}\left(\frac{d^+}{dt}+\frac{d^-}{dt}\right)+i\frac{\eta}{2}\left(\frac{d^+}{dt}-\frac{d^-}{dt}\right),
\end{equation}
and
\begin{equation}
\begin{aligned}
\lambda_{k_1,\dots,k_{j_\alpha}}=&\frac{1}{2}\left(\lambda^+_{k_1} \dots \lambda^+_{k_{j_\alpha}}+(-1)^{j_\alpha-1}\lambda^-_{k_1} \dots \lambda^-_{k_{j_\alpha}}\right)\\
&+i\frac{\eta}{2}\left(\lambda^+_{k_1} \dots \lambda^+_{k_{j_\alpha}}+(-1)^{j_\alpha}\lambda^-_{k_1} \dots \lambda^-_{k_{j_\alpha}}\right).
\end{aligned}
\end{equation}

\section{Asymptotic Newton's equation under fractional scale regime}
\label{asymptotic}
We consider the classical equation obtained by Newton to describe the dynamical behavior of a particle of mass $m$ under the action of a force deriving from a potential $U$. Precisely, we call {\it Newton's equation} the following ordinary differential equation
\begin{equation}
\di m\frac{d \V}{dt} = -\grad{U(\X)},
\end{equation}
with $\X\in \R^d$ and where $\V=\frac{d\X}{dt}$. This equation can be seen as a result of a scale formulation with a linear scale regime, i.e. the asymptotic behavior is differentiable. We now consider the scale formulation of the Newton equation under a fractional scale regime. The asymptotic Newton's equation associated is given by (see \cite{cresson-pierret_scale})
\begin{equation}
m\Boxinf \V_\infty= -\grad{U(\X_\infty)}
\end{equation}
where $\V_\infty = \Boxinf \X_\infty := \frac{\Box}{\Box t} \X^\star_\infty$. As there is no confusion possible with \emph{scale functions} introduced in \cite{cresson-pierret_scale}, for notation convenience, we remove the $\infty$ sign on $\X$ and $\V$. \\

The Lagrangian formulation of the asymptotic Newton's equation allows us to relate the velocity to the action functional $\mathcal{A}(t,\X)$ as
\begin{equation}
m\V(t,\X) = \grad{\mathcal{A}(t,\X)}
\end{equation}
where the function $\mathcal{A}(t,\mathbf{X})$ is differentiable with respect to t and $\mathbf{X}$. From this formulation, we obtain the following \emph{asymptotic fractional Hamilton-Jacobi equation}:
\begin{lemma}
\label{lemma_hj}
The asymptotic fractional Hamilton-Jacobi equation associated to the action functional $\mathcal{A}$ is given by
\begin{equation}
	\frac{\partial\mathcal{A}}{\partial t}+\frac{\left(\grad{\mathcal{A}}\right)^2}{2m}+\sum_{k_1,\dots,k_{j_\alpha}=1}^{d}\frac{\lambda_{k_1 \dots k_{j_\alpha}}}{{j_\alpha}!}\,\frac{\partial^{j_\alpha} \mathcal{A}}{\partial x_{k_1} \dots \partial x_{k_{j_\alpha}}}+U=0
\end{equation}
\end{lemma}
The proof is given in Appendix \ref{dem_lemma_hj}.\\

The complex definition of the Box derivative $\frac{\Box}{\Box t}$ induces two components $(\textbf{v},\textbf{u})$ for the velocity as $\V = \textbf{v}+i\eta \textbf{u}$. In consequence, we can decompose the action functional $\mathcal{A}$ in two parts $(\mathcal{S},\mathcal{R})$ as follows
\begin{equation}
\mathcal{A}= \mathcal{S}+i\eta \mathcal{R},
\end{equation}
with $\textbf{v}=\frac{\grad{\mathcal{S}}}{m}$ and $\textbf{u}=\frac{\grad{\mathcal{R}}}{m}$. Now, identifying the real and the imaginary parts of the asymptotic fractional Hamilton-Jacobi equation, we obtain
\begin{corollary}
Let $\lambda=\lambda_\Re +i\eta \lambda_\Im$. The real and imaginary parts of the asymptotic fractional Hamilton-Jacobi equation are respectively given by
\begin{equation}
\label{hj-system}
\left\{\begin{aligned}
&\frac{\partial \mathcal{S}}{\partial t} +\frac{\left(\grad{\mathcal{S}}\right)^2}{2m}-\eta^2\frac{\left(\grad{\mathcal{R}}\right)^2}{2m}+\sum_{k_1,\dots,k_{j_\alpha}=1}^{d}\frac{1}{{j_\alpha}!}\,\frac{\partial^{j_\alpha} }{\partial x_{k_1} \dots \partial  x_{k_{j_\alpha}}}\left(\lambda_\Re\mathcal{S}-\eta^2\lambda_\Im\mathcal{R}\right)+U=0,\\
&\frac{\partial \mathcal{R}}{\partial t} + \frac{\grad{\mathcal{S}}\cdot\grad{\mathcal{R}}}{m}+\sum_{k_1,\dots,k_{j_\alpha}=1}^{d}\frac{1}{{j_\alpha}!}\,\frac{\partial^{j_\alpha} }{\partial x_{k_1} \dots \partial x_{k_{j_\alpha}}}\left(\frac{\lambda_\Im}{2}\mathcal{S}+\frac{\lambda_\Re}{2}\mathcal{R}\right)=0.
\end{aligned}\right.
\end{equation}
\end{corollary}

In the linear scale regime case, i.e. $\lambda_{k_1 \dots k_{j_\alpha}}=0$ for all $k_1 \dots k_{j_\alpha}\geq 1$ and $\mathcal{R}=0$, the system of equations \eqref{hj-system} reduces to the usual Hamilton-Jacobi equation
\begin{equation}
\frac{\partial \mathcal{S}}{\partial t} +\frac{\left(\grad{\mathcal{S}}\right)^2}{2m}+U=0,
\end{equation}
where the Hamiltonian $H_\mathcal{S}$ associated to the action functional $\mathcal{S}$ is defined as $H_\mathcal{S}=-\frac{\partial \mathcal{S}}{\partial t}$. In the fractional scale regime it defines a two dimensional Hamiltonian $\textbf{H}_{\mathcal{S},\mathcal{R}}$ as
\begin{equation}
\textbf{H}_{\mathcal{S},\mathcal{R}}=-\frac{\partial}{\partial t}\left(\begin{matrix}\mathcal{S} \\ \mathcal{R}\end{matrix}\right).
\end{equation}
As we can see, the first component of this two dimensional Hamiltonian can be seen as \emph{adding} extra terms to the Hamiltonian of the linear scale regime $H_\mathcal{S}$. Knowing the solution of $\mathcal{R}$, it can be interpreted as a \emph{modification} or \emph{an addition} of the potential $U$. This is exactly the motivation of using the framework of \emph{scale dynamics} which induces naturally dynamical effects depending on the scale regime. \\

In order to illustrate our formalism and to apply it to mechanics, we made the following assumptions on the fractional scale regime:

%
%
%

\begin{assumption}
\label{assum1}
The fractional scale regime is of order $\alpha=\frac{1}{2}$, i.e. ${j_\alpha}=2$.
\end{assumption}

\begin{assumption}
The fractional scale regime is uniform and has independent components, i.e. for all $1\leq i,j\leq d$ and $i\neq j$ we have \begin{equation*}
	\lambda^+_{k_i} \lambda^+_{k_j} = \left(\lambda^+\right)^2\delta_{k_i,k_j}\ \text{and}\ \lambda^-_{k_i} \lambda^-_{k_j} = \left(\lambda^-\right)^2\delta_{k_i,k_j},
	\end{equation*}
	where $\delta$ is the Kronecker delta. It follows that $\lambda_{k_i k_j} =\lambda\delta_{k_i,k_j}$ with
	\begin{equation*}
	\lambda = \left(\frac{\lambda^+-\lambda^-}{2}+i\eta \frac{\lambda^++\lambda^-}{2}\right).
	\end{equation*}
\end{assumption}
In consequence, we obtain the following lemma:
\begin{lemma}
Under Assumptions I and II, the asymptotic fractional Hamilton-Jacobi equation of order $2$ associated to the action functional $\mathcal{A}$ is given by
\begin{equation}
\label{hj-2}
\frac{\partial\mathcal{A}}{\partial t}+\frac{\left(\grad{\mathcal{A}}\right)^2}{2m}+\frac{\lambda}{2}\,\lap{\mathcal{A}}+U=0
\end{equation}
and its real and imaginary parts are given by
\begin{equation}
\label{hj-system2}
\left\{\begin{aligned}
&\frac{\partial \mathcal{S}}{\partial t} +\frac{\left(\grad{\mathcal{S}}\right)^2}{2m}+\frac{\lambda_\Re}{2}\lap{\mathcal{S}}-\eta^2\left(\frac{\left(\grad{\mathcal{R}}\right)^2}{2m}+\frac{\lambda_\Im}{2}\lap{\mathcal{R}}\right)+U=0,\\
&\frac{\partial \mathcal{R}}{\partial t} + \frac{\grad{\mathcal{S}}\cdot\grad{\mathcal{R}}}{m}+\frac{\lambda_\Im}{2}\lap{\mathcal{S}}+\frac{\lambda_\Re}{2}\lap{\mathcal{R}}=0.
\end{aligned}\right.
\end{equation}
\end{lemma}

In classical mechanics, an important relation can be derived from the Newton equation which is known as the \emph{Virial theorem}. The asymptotic fractional Newton equation allows obtaining a generalization of this theorem.

\begin{lemma}
\label{virial}
Let $I=m\X^2$ be the quantity called \emph{the moment of inertia}. Under Assumptions I and II and assuming the potential $U$ is a homogeneous function of order $\gamma$ then, we have
\begin{equation}
\frac{1}{2}\Boxinf^2 I = 2K-\gamma U + \lambda m\div{\V},
\end{equation}
where $K=\frac{1}{2}m\V^2$ is the kinetic energy. If the system is at the equilibrium, i.e. $\frac{1}{2}\Boxinf^2 I=0$ then, we have the generalized Virial theorem given by the relation
\begin{equation}
\label{virial_eq}
2K + \lambda m\div{\V}=\gamma U.
\end{equation}
\end{lemma}
The proof is given in Appendix \ref{dem_virial}.

\section{From Newton to Schr\"odinger equation and vice versa}

In order to obtain $\mathcal{R}$ to have the induced dynamical effects from the scale regime on the classical motion, a way to solve analytically the asymptotic fractional Hamilton-Jacobi equation \eqref{hj-2} or the system \eqref{hj-system2}, is to use the following change of variable:
\begin{equation}
\label{psi}
\psi(t,\X)=e^{\frac{-\eta \mathcal{R}(t,\X)+i\mathcal{S}(t,\X)}{K}}
\end{equation}
with $K$ a real constant. It follows that $\mathcal{A}(t,\X)=-iK\ln \psi(t,\X)$. Using the same kind of computations as in \cite{cresson-pierret_scale}, from the asymptotic fractional Hamilton-Jacobi equation \eqref{hj-2}, we obtain the following partial differential equation satisfied by $\psi$:
\begin{equation}
\label{schro}
	iK\frac{\partial \psi}{\partial t}+\frac{iK\lambda}{2}\lap{\psi}+\frac{\left(\grad{\psi}\right)^2}{\psi}\left(\frac{K}{m}-i\lambda\right)\frac{K}{2}-U\psi =0,
\end{equation}
which is the \emph{non-linear Schr\"odinger equation}. A convenient way to write the function $\psi$ is to consider the positive defined function $P$ as $\sqrt{P}=e^{-\frac{\eta\mathcal{R}}{K}}$. In that case, we have $\psi = \sqrt{P}e^\frac{i\mathcal{S}}{K}$ and from this definition, we obtain the following lemma:

\begin{lemma}
Considering the change of variable \eqref{psi} with $\sqrt{P}=e^{-\frac{\eta\mathcal{R}}{K}}$, the asymptotic fractional Hamilton-Jacobi equation \eqref{hj-2} is equivalent to
\begin{equation}
\label{hj-system3}
\resizebox{\textwidth}{!}{$\left\{\begin{aligned}
&\frac{\partial \mathcal{S}}{\partial t} +\frac{\left(\grad{\mathcal{S}}\right)^2}{2m}+\frac{\lambda_\Re}{2}\lap{\mathcal{S}}-\frac{K^2}{2m}\frac{\lap{(\sqrt{P})}}{\sqrt{P}}+\frac{K}{2}\left(\frac{K}{m}+\eta\lambda_\Im\right)\lap{\left(\ln\sqrt{P}\right)}+U=0,\\
&\frac{\partial P}{\partial t} + \div{\left(P\cdot\frac{\grad{S}}{m}\right)}-P\cdot\frac{\lap{S}}{m}\left(1+\eta\frac{ m\lambda_\Im}{K}\right)+\frac{K\lambda_\Re}{2}\lap{(\ln\sqrt{P})}=0
\end{aligned}\right.$}
\end{equation}
\end{lemma}
The proof is given in Appendix \ref{dem_hj-system3}.\\

From the previous derivation, we have a general formulation of the asymptotic Newton equation through the asymptotic fractional Hamilton-Jacobi equation and the Schr\"odinger equation under Assumptions I and II. In order to solve the non-linear Schr\"odinger equation \eqref{schro} for the Kepler problem in the next section, we make a last assumption on the fractional scale regime:
\begin{assumption}
The fractional scale regime is uniform over time derivatives, i.e. $\Lambda:=\lambda^+=\lambda^-$.
\end{assumption}

Then, we have:
\begin{corollary}
\label{forme_finale}
Specializing the $\Boxinf$ derivative to $\eta=-1$, under Assumptions I--III, the \emph{non-linear Schr\"odinger equation} \eqref{schro} is equivalent
\begin{equation}
\label{schro_nl}
iK\frac{\partial \psi}{\partial t}+\frac{K\Lambda}{2}\lap{\psi}+\frac{\left(\grad{\psi}\right)^2}{\psi}\left(\frac{K}{m}-\Lambda\right)\frac{K}{2}-U\psi =0,
\end{equation}
and the asymptotic fractional Hamilton-Jacobi equation \eqref{hj-system3} is equivalent to
\begin{equation}
\left\{\begin{aligned}
&\frac{\partial \mathcal{S}}{\partial t} +\frac{\left(\grad{\mathcal{S}}\right)^2}{2m}-\frac{K^2}{2m}\frac{\lap{(\sqrt{P})}}{\sqrt{P}}+\frac{K}{2}\left(\frac{K}{m}-\Lambda\right)\lap{\left(\ln\sqrt{P}\right)}+U=0,\\
&\frac{\partial P}{\partial t} + \div{\left(P\cdot\frac{\grad{S}}{m}\right)}+P\cdot\frac{\lap{S}}{m}\left(\frac{ m\Lambda}{K}-1\right)=0
\end{aligned}\right.
\end{equation}

In the special case where $K=m\Lambda$ the \emph{non-linear Schr\"odinger equation} \eqref{schro} is equivalent to the \emph{linear Schr\"odinger equation}
\begin{equation}
\label{schro_l}
im\Lambda\frac{\partial \psi}{\partial t}+\frac{m\Lambda^2}{2}\lap{\psi}-U\psi =0,
\end{equation}
and the asymptotic fractional Hamilton-Jacobi equation \eqref{hj-system3} is equivalent to
\begin{equation}
\label{last}
\left\{\begin{aligned}
&\frac{\partial \mathcal{S}}{\partial t} +\frac{\left(\grad{\mathcal{S}}\right)^2}{2m}-\frac{m\Lambda^2}{2}\frac{\lap{(\sqrt{P})}}{\sqrt{P}}+U=0,\\
&\frac{\partial P}{\partial t} + \div{\left(P\cdot\frac{\grad{S}}{m}\right)}=0.
\end{aligned}\right.
\end{equation}
\end{corollary}

\begin{remark}
The extra term $-\frac{m\Lambda^2}{2}\frac{\lap{(\sqrt{P})}}{\sqrt{P}}$ in the first equation of the asymptotic fractional Hamilton-Jacobi equation lead us to interpret it as an additional potential which found its nature in the fractional scale regime of the motion.
\end{remark}

\section{Application to the Kepler problem}
\label{kepler}
We now apply our previous derivation to the Kepler problem. Consider two bodies of mass $M$ and $m$ in the Euclidean space $\R^3$. In that case, the potential $U$ is a homogeneous function or order $-1$ and is given by the well-known relation
\begin{equation}
U=-\frac{k}{r},
\end{equation}
with $k=GMm$, $G$ is the universal constant of gravitation and $r=\sqrt{\X \cdot \X}$ is the distance between the two bodies at each instant. \\

In order to solve the Schr\"odinger equations \eqref{schro_nl} and \eqref{schro_l}, we use the separation variable method. Indeed, we look for function $\psi$ as
\begin{equation}
\psi(t,\X) = f(t) \Psi(\X),
\end{equation}
with $f(t)=e^{-\frac{iE t}{K}}$, $E$ the total energy of the system and $\Psi$ is a function depending only on the position. In consequence, the \eqref{schro_nl} is equivalent to
\begin{equation}
\label{schro_ind}
\lap{\Psi}+\frac{\left(\grad{\Psi}\right)^2}{\Psi}\left(\frac{K}{m\Lambda}-1\right)+\frac{2}{K\Lambda}\left(E-U\right)\Psi =0,
\end{equation}
In spherical coordinates $(r,\phi,\theta)$, the function $\Psi$ can be expressed in terms of three functions $R,\Phi,\Theta$ as
\begin{equation}
\Psi(\X)=R(r)\Theta(\theta)\Phi(\phi).
\end{equation}
Inserting the expression of $\Psi(\X)$ in the Equation \eqref{schro_ind} lead us to solve by the separation variable method, the three differential equations:
\begin{equation}
\label{separation}
\left\{\begin{aligned}
&\frac{d^2 R}{dr^2}+\frac{2}{r}\frac{dR}{dr}+\left(\frac{K}{m\Lambda}-1\right)\frac{1}{R}\left(\frac{dR}{dr}\right)^2+\left(\frac{2}{K\Lambda}(E-U)-\frac{C'}{r^2}\right)R=0,\\
& \frac{d^2\Theta}{d\theta^2}+\frac{1}{\tan \theta }\frac{d\Theta}{d\theta}+\left(\frac{K}{m\Lambda}-1\right)\frac{1}{\Theta}\left(\frac{d\Theta}{d\theta}\right)^2+\left(C'-\frac{C}{\sin^2\theta}\right)\Theta=0 \\
&\frac{d^2\Phi}{d\phi^2}+\left(\frac{K}{m\Lambda}-1\right)\frac{1}{\Phi}\left( \frac{d\Phi}{d\phi}\right)^2-C\Phi=0.
\end{aligned}\right.
\end{equation}
where $C$ and $C'$ are the real constants from the successive use of separation variable method. Now, we look for the so-called \emph{ground state solution} of the Schrodinger equation, i.e. the constants $C$ and $C'$ are set to zero. The ground state energy $E=-E_0$ is defined by $E_0=\frac{k^2}{2m^2\Lambda^2}$.\\

In the linear case, the ground state solution of the Schr\"odinger equation \eqref{schro_l} corresponds to the well-known ground state solution of the Hydrogen atom model (see \cite{cohen1977quantum}) but in this problem, with the Keplerian potential. Solving Equations \eqref{separation}, the solution in term of $\sqrt{P}$ is given by
\begin{equation}
\sqrt{P}=\left(\frac{C_1+C_2}{m\Lambda^2}\right)e^{-\frac{2r}{r_0}},
\end{equation}
where $r_0=\frac{2\Lambda^2}{G M}$ and $C_1$, $C_2$ are two integration real constants.

\begin{remark}
We keep the arbitrary constants of integration because we do not need their explicit value for what follows.
\end{remark}
In consequence, we obtain the expression of extra term denoted $U_{\text{add}}=-\frac{m\Lambda^2}{2}\frac{\lap{(\sqrt{P})}}{\sqrt{P}}$ as
\begin{equation}
U_\text{add}=-\frac{GMm}{r_0}\left(1-\frac{r_0}{r}\right).
\end{equation}
In the non-linear case, we obtain the \emph{ground state solution} in term of $\sqrt{P}$ as
\begin{equation}
\sqrt{P}=C_1 e^{-\frac{GM m }{\Lambda  K}r-\frac{\Lambda  m }{K}\ln r+\frac{m\Lambda}{K}\ln \left(\Lambda ^2 e^{\frac{2 r}{r_0}}-2 GM \textsf{Ei}\left(\frac{2r}{r_0}\right)r-C_2 \Lambda ^2 r\right)}
\end{equation}
where $C_1$, $C_2$ are two real integration constants and $\textsf{Ei}$ is the exponential integral function defined by $\textsf{Ei}(x)=\int_{-\infty}^{x}\frac{e^{-t}}{t}dt$. In that case, we obtain the same expression for $U_\text{add}$. \\

\begin{remark}
We expected to obtain the same extra potential in the two cases. Indeed, the non-linearity is linked to the choice of the constant $K$. This constant could not be independent of the problem and induces a dynamical effect because it would add an arbitrary degree of freedom. Even if, in the change of variable with the $\psi$ function, $K$ could be chosen arbitrary, the contribution of the non-linearity related to $K$ and its arbitrary choice is destroyed in the asymptotic fractional Hamilton-Jacobi.
\end{remark}

\begin{figure}
	\centering
	\includegraphics[width=0.8\textwidth]{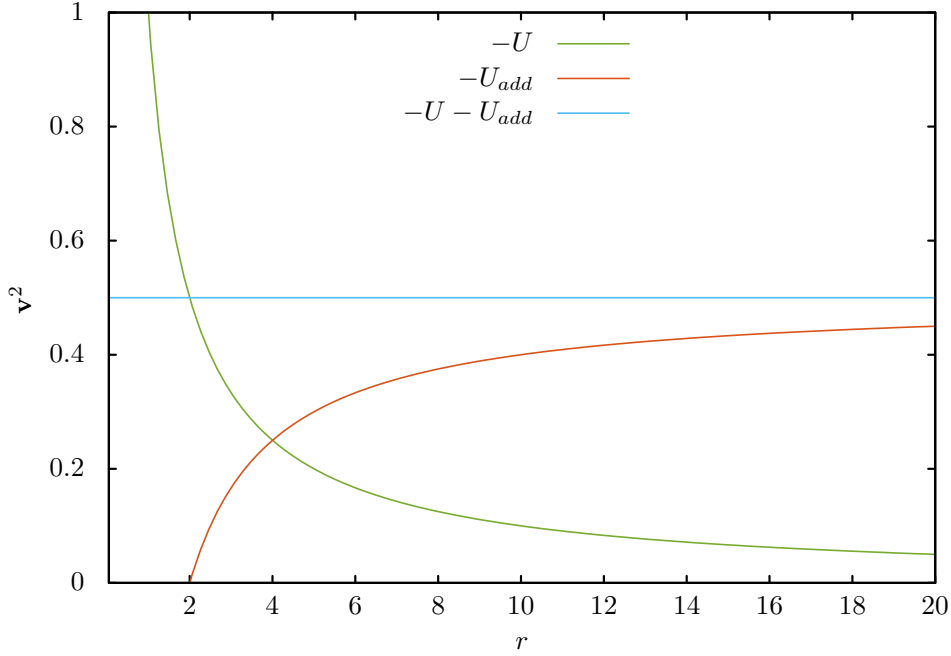}
	\caption{Illustration of the square of the orbiting body velocity depending of the potential term in the fractional Virial theorem.}
	\label{velocity_curve}
\end{figure}

In order to obtain of the velocity of the orbiting body, we use our derivation of the generalized Virial theorem \eqref{virial_eq}. The real part of Equation \eqref{virial_eq} is equivalent to
\begin{equation}
m\textbf{v}^2=-U+\frac{m\Lambda^2}{2}\frac{\lap{(\sqrt{P})}}{\sqrt{P}}=-U-U_{add}.
\end{equation}
In consequence, we obtain
\begin{equation}
\|\textbf{v}\|=\sqrt{\frac{GM}{r_0}},
\end{equation}
which means that the velocity of the orbiting body is constant (see Figure \ref{velocity_curve} for an illustration with $GM=1$ and $\Lambda=1$).
\begin{remark}
In this particular case of fractional scale regime, we recover rigorously the result obtained in \cite{nottale2011, rocha2003}.
\end{remark}

Such a situation of Keplerian motion would appear in the outer region of a galaxy. It would explain the observed constant velocity in spiral galaxies far away from its galactic center. In consequence, modifying the law of gravitation as in MOND theory would only result in interpolating the supplementary terms appearing due to the scale formulation of Newton's equation, which is in that case, a scale formulation with a fractional scale regime.

\section{Conclusion}

In this paper, we showed that the scale dynamical formulation of Newton's equation generates  supplementary dynamical terms which could be interpret as a modification of the potential, as in MOND theory, or an addition of an extra potential, as it is done with dark matter. Moreover, it is supported by the intriguing fact that when considering our approach for the two body problem, it shows that the velocity of the orbiting body tends to be a constant, as it is observed for the flat rotation curves of galaxies. One has to notice that more complicated scale regime can be chosen such as logarithmic comparison scale or Hardy comparison scale (see \cite{tricot}) but as we can see, choosing the fractional scale regime of order $1/2$ already leads to a probable explanation of strange effects such as the dark matter.

\section{Appendix}

\subsection{Proof of Lemma \ref{lemma_hj}}
\label{dem_lemma_hj}
By definition of the asymptotic Box derivative $\Boxinf$, we have
\begin{equation}
\begin{aligned}
\Boxinf\left(\grad{\mathcal{A}(t,\X)}\right)=&\frac{\partial \grad{\mathcal{A}(t,{\X^\star})}}{\partial t} + \grad{\left(\grad{\mathcal{A}(t,{\X^\star})}\right)}\cdot \Boxinf {\X^\star} \\ &+\sum_{k_1,\dots,k_{j_\alpha}=1}^{d}\frac{\lambda_{k_1 \dots k_{j_\alpha}}}{{j_\alpha}!}\,\frac{\partial^{j_\alpha} \grad{\mathcal{A}(t,{\X^\star})}}{\partial x_{k_1} \dots \partial x_{k_{j_\alpha}}}.
\end{aligned}
\end{equation}
As $\Boxinf \X = \V = \frac{\grad{\mathcal{A}}}{m}$ and $\mathcal{A}$ is differentiable, we obtain
\begin{align}
\frac{\partial \grad{\mathcal{A}(t,{\X^\star})}}{\partial t} & = \grad{\frac{\partial \mathcal{A}(t,{\X^\star})}{\partial t}}, \\
\grad{\left(\grad{\mathcal{A}(t,{\X^\star})}\right)}\cdot \Boxinf {\X^\star} & = \grad{\left(\frac{\left(\grad{\mathcal{A}(t,{\X^\star})}\right)^2}{2m}\right)}.
\end{align}
As the partial derivatives and the gradient commute for the scalar function $\mathcal{A}$, we obtain
\begin{equation}
\begin{aligned}
\Boxinf\left(\grad{\mathcal{A}(t,\X)}\right)= \mathsf{grad}&\left[\frac{\partial {\mathcal{A}(t,{\X^\star})}}{\partial t} + \frac{\left(\grad{\mathcal{A}(t,{\X^\star})}\right)^2}{2m} \right. \\
&+\left. \sum_{k_1,\dots,k_{j_\alpha}=1}^{d}\frac{\lambda_{k_1 \dots k_{j_\alpha}}}{{j_\alpha}!}\,\frac{\partial^{j_\alpha} \mathcal{A}(t,{\X^\star})}{\partial x_{k_1} \dots \partial x_{k_{j_\alpha}}}\right].
\end{aligned}
\end{equation}
From the asymptotic Newton equation, we obtain
\begin{equation}
\begin{aligned}
\mathsf{grad}&\left[\frac{\partial {\mathcal{A}(t,{\X^\star})}}{\partial t} + \frac{\left(\grad{\mathcal{A}(t,{\X^\star})}\right)^2}{2m} \right. \\
&+ \left.\sum_{k_1,\dots,k_{j_\alpha}=1}^{d}\frac{\lambda_{k_1 \dots k_{j_\alpha}}}{{j_\alpha}!}\,\frac{\partial^{j_\alpha} \mathcal{A}(t,{\X^\star})}{\partial x_{k_1} \dots \partial x_{k_{j_\alpha}}}+U\right]=0.
\end{aligned}
\end{equation}
Integrating with respect to the spatial variable, we obtain
\begin{equation}
\frac{\partial {\mathcal{A}(t,\X^\star)}}{\partial t} + \frac{\left(\grad{\mathcal{A}(t,{\X^\star})}\right)^2}{2m} + \sum_{k_1,\dots,k_{j_\alpha}=1}^{d}\frac{\lambda_{k_1 \dots k_{j_\alpha}}}{{j_\alpha}!}\,\frac{\partial^{j_\alpha} \mathcal{A}(t,{\X^\star})}{\partial x_{k_1} \dots \partial x_{k_{j_\alpha}}}+U=C({\X^\star}),
\end{equation}
where $C({\X^\star})$ is an arbitrary continuous function. We can always choose $C({\X^\star})=0$ and by consequence, we obtain the result.

\subsection{Proof of Lemma \ref{virial}}
\label{dem_virial}

We have
\begin{equation}
\Boxinf I = 2m\left(\X \cdot \frac{\Box}{\Box t}\X^\star + \lap{\left(\X^\star\right)^2}\right).
\end{equation}
By definition, $\Boxinf \X = \frac{\Box}{\Box t} \X^\star=\V$ then,
\begin{equation}
\Boxinf^2 I = 2m\left[\frac{\Box}{\Box t} \X^\star \cdot \Boxinf \X + \X \cdot \frac{\Box}{\Box t} \left(\Boxinf \X\right) + \frac{\lambda}{2} \lap{(\Boxinf \X \cdot \X)}\right].
\end{equation}
As $\lap{(\V \cdot \X)}=(\lap{\V})\cdot\X+2\div{\V}$. Then, we have
\begin{equation}
\Boxinf^2 I = 2m \V \cdot \V + 2m\X \cdot \left[\frac{\Box}{\Box t} \V + \frac{\lambda}{2}\lap{\V}\right] +2m\lambda\div{\V}.
\end{equation}
By definition of $\Boxinf$ and using the asymptotic Newton equation, we obtain
\begin{equation}
2m\X \cdot \left[\frac{\Box}{\Box t} \V + \frac{\lambda}{2}\lap{\V}\right]=-2 \X \cdot \grad{U}.
\end{equation}
As $U$ is an homogeneous function of order $\gamma$, we have $\X \cdot \grad{U}=\gamma U$. Inserting this expression into $\Boxinf^2 I$ concludes the proof.

\subsection{Proof of Lemma \ref{hj-system3}}
\label{dem_hj-system3}

Remarking that for any function $f$, we have the identity
\begin{equation}
\left(\grad{(\ln f)}\right)^2+\lap{\left(\ln f\right)}=\frac{\lap f}{f}.
\end{equation}
Then, by definition and using the identity with $f=\sqrt{P}$, we obtain
\begin{equation}
\frac{\left(\grad{\mathcal{R}}\right)^2}{2m}+\frac{\lambda_\Im}{2}\lap{\mathcal{R}} =\frac{K^2}{2m}\frac{\lap{(\sqrt{P})}}{\sqrt{P}}+\frac{K}{2}\left(\frac{K}{m}+\eta \lambda_\Im\right)\lap{\left(\ln\sqrt{P}\right)}.
\end{equation}
The last steps of the proof follows from simple computations.

\bibliographystyle{alpha}

\end{document}